\documentclass[12pt]{article}
\RequirePackage{ifpdf}
\ifpdf\RequirePackage[raiselinks=false,colorlinks=true,citecolor=blue,urlcolor=blue,linkcolor=blue,bookmarksopen=true,pdftex]{hyperref}\else
\RequirePackage[raiselinks=false,colorlinks=true,citecolor=blue,urlcolor=blue,linkcolor=blue,bookmarksopen=true,dvips]{hyperref}\fi
\usepackage[latin1]{inputenc} 
\usepackage[T1]{fontenc} 
\usepackage{amsfonts}
\usepackage{latexsym}
\usepackage{color}  %
\usepackage{ae,aecompl,amsbsy,amssymb}
\usepackage{epsf}
\usepackage{graphics}
\usepackage{framed}
\usepackage{combelow}

\newcommand{\limp}{\Rightarrow}

\newcommand{\Rocq}{{\sc Rocq}}
\newcommand{\Ssreflect}{{\sc SSReflect}}
\usepackage{tikz}
\usetikzlibrary{arrows.meta,calc,decorations.markings,math,arrows.meta}
\newlength{\hsbw}
\newcommand\MSpacing{13pt}

\title{Revisiting the Fast Fourier Transform in Rocq}
\author{Laurent Th{\'e}ry \\
Stamp Team - INRIA \\
{\tt Laurent.Thery@inria.fr}}
\date{}
\begin{document}
\maketitle
\begin{abstract}

This note explains how a standard algorithm that constructs
the discrete Fourier transform has been formalised and proved correct 
in  the 
{\Rocq} proof assistant using the {\Ssreflect} extension.
\end{abstract}

\section{Introduction}

Fast Fourier Transforms are key tools in many areas. In this note,
we are going to explain how they have been formalised in a theorem 
prover like {\Rocq}. The presentation of this work has three parts.
The central part of this work is the formalisation of the algorithm in itself.
We have aimed to get a nice and effective formal presentation. 
Then, two extensions have been developed. 
The first one builds a dedicated tactic using the Fourier algorithm in order
to implement polynomial multiplication. 
The second one goes toward computer arithmetic and formally  proves some standard 
results about error bounds while computing Fourier transform with 
floating-point numbers. 

Note that this is not the first time this algorithm has been 
formalised in {\Rocq}. To our knowledge, the first one was done in 2001 by Venanzio 
Capretta (see~\cite{capretta}).
It is about time to revisit this work and see how concise 
it can get using existing libraries. Also, this initial effort was 
mostly interested
in the recursive presentation of the algorithm. Here, we also give an  
iterative version.

The paper is organised as follows. We first describe the formalisation 
of the main algorithm. Then, we first present the tactic that has been  
built on top of the algorithm. Finally, we mention some results concerning 
error bounds when computing with floating-point numbers.

\section{Main algorithms}

\subsection*{The recursive algorithm}

The algorithm manipulates three kinds of data:
\begin{itemize}
\setlength\itemsep{-0.2em}
\item[-] natural numbers, i.e. elements of type \texttt{nat};
\item[-] elements of an arbitrary integral domain $R$;
\item[-] univariate polynomials over $R$, i.e. elements of type \texttt{\{poly $R$\}}.
\end{itemize}
\noindent
The operations we use on natural numbers are the successor and predecessor functions 
($i$\texttt{.+1} and $i$\texttt{.-1}),
the doubling and halving functions ($i$\texttt{.*2}, $i$\texttt{./2} and 
\texttt{uphalf} $i$ $=$ ($i$\-\texttt{.+1}\-\texttt{./2})),
the exponentiation ($i$\,\texttt{\^}\,$j$),
the division ($i$\,\,\texttt{\%/}\,\,$j$),
 and the modulo ($i$\,\,\texttt{\%\%}\,\,$j$).
For the integral domain, we use the usual ring operations : the addition $x$ \texttt{+} $y$, 
the multiplication by a scalar  $k$ \texttt{*:} $x$, the multiplication 
$x$ \texttt{*} $y$, and the exponentiation by a natural number $x$ \texttt{\^}\texttt{+} $n$.
Also, a predicate of the library that is useful in our application is the one that indicates that 
an element $w$ is a $n^\textit{th}$ primitive root of unity. It is written 
as $n$\texttt{.-primitive\_root} $w$. It means that $w$ \texttt{\^}\texttt{+} $n$ $=$ $1$
and that $n$ is actually the smallest non-zero natural number that has this property.
In our formalisation, we derive two easy lemmas about primitive roots
\noindent
\vskip0pt
\noindent
\begin{framed}
\footnotesize
\noindent
\vskip0pt\noindent\hskip0pt
\texttt{Lemma} \textit{prim\_exp2nS} $n$ ($w$ : $R$) :
\vskip0pt\noindent\hskip10pt
(2 \texttt{\^} $n$\texttt{.+1})\texttt{.-primitive\_root} $w$ \texttt{$\rightarrow$} $w$ \texttt{\^}\texttt{+} (2 \texttt{\^} $n$) \texttt{=} -1.
\vskip0pt\noindent\hskip0pt
\texttt{Lemma} \textit{prim\_sqr} $n$ ($w$ : $R$) : 
\vskip0pt\noindent\hskip10pt
(2 \texttt{\^} $n$\texttt{.+1})\texttt{.-primitive\_root} $w$ \texttt{$\rightarrow$} (2 \texttt{\^} $n$).\texttt{-primitive\_root} ($w$ \texttt{\^}\texttt{+} 2).
\end{framed}
\vskip0pt
\noindent
The first lemma is used to simplify expression involving $w^j$. Its proof is the only 
place where the integrality is used (in order to get from $w^2 =1$
that either $w = 1$ or $w = -1$). 
The recursive algorithm of the fast Fourier transform takes as argument a primitive 
root $w$ and performs some recursive calls with $w$ \texttt{\^}\texttt{+} 2. The
lemma \textit{prim\_sqr} is then used to prove that the primitive root property of the argument 
is an invariant of the recursion. 

The algorithm is mainly manipulating univariate polynomials. A polynomial 
is represented by a list whose last element (the leading term), if it exists, is 
non-zero. The empty list represents the null polynomial. A polynomial $p$ 
can be automatically converted to a list, so \texttt{size} $p$ is understood 
as the length of the list representing $p$. So, if $p$ is not null, it is 
the usual degree of the polynomial incremented by one. We can access the $n^\textit{th}$
term of polynomial $p$ by $p$\texttt{\textquotesingle\_}$n$.
The leading term is then $p$\texttt{\textquotesingle\_} (\texttt{size} $p$)\texttt{.-1}.
The unique variable of the univariate polynomials is written \textquotesingle\texttt{X}
and \textquotesingle\texttt{X\^}$n$ its power.
Evaluating a polynomial $p$ at point $x$ is written $p$\texttt{.[}$x$\texttt{]}.
Composition two polynomials $p$ and $q$ which consists in lifting the 
evaluation from points to polynomials  $p$\texttt{.[}$q$\texttt{]}
is written $p$ \texttt{$\backslash$Po} $q$. 
We can turn a function $F$ into a polynomial using 
 \texttt{{$\backslash$}poly\_}($i$ \texttt{<} n) $F$ $i$ that builds a polynomial
 of size at most $n$ whose $i^\textit{th}$ term is $F$ $i$. A special notation is available
 for constant polynomials where one can write $c$\texttt{\%:P} in order to build  
 a polynomial that only contains the constant term $c$. 
 
Our recursive algorithm is taking two arguments : 
a polynomial $p$ and a primitive root $w$ of degree $2 ^ n$ and 
returns the evaluation of $p$ at points 1, $w$, $w^2$, $\dots$, $w^{2^n -1}$ as 
 the polynomial $p$\texttt{.[}$1$\texttt{]} \texttt{+} $p$\texttt{.[}$w$\texttt{]} $X$ 
 \texttt{+} \dots \texttt{+} $p$\texttt{.[}$w^{2^n -1}$\texttt{]} $X^{2^n -1}$. 
So, the return type is also a polynomial.
 The recursive
 calls are made on the even and odd parts of the polynomial $p$.
If $p = 1 + 2 X + 3X^2 + 4X^3 + 5 X^4$, its even part is $1 + 3X + 5 X^2$
and its even part  $2 + 4 X$.
These operations on polynomials are in the library.
\noindent
\vskip0pt
\noindent
\begin{framed}
\footnotesize
\noindent
\vskip0pt\noindent\hskip0pt
\texttt{Definition} \textit{even\_poly} $p$ \texttt{:} \texttt{\{poly $R$\}} \texttt{:=}
\texttt{$\backslash$poly\_}($i$ < \texttt{uphalf} (\texttt{size} $p$)) $p$\texttt{\textquotesingle\_}$i$.\texttt{*2}.
\vskip0pt\noindent\hskip0pt
\texttt{Definition} \textit{odd\_poly}\,\,\,\,\,$p$ \texttt{:} \texttt{\{poly $R$\}} \texttt{:=} \texttt{$\backslash$poly\_}($i$ $<$ (\texttt{size} $p$)\texttt{./2}) $p$\texttt{\textquotesingle\_}$i$\texttt{.*2.+1}.
\end{framed}
\vskip0pt
\noindent
There is also the key lemma that justifies the decomposition of 
the polynomial in the recursive calls 
\noindent
\vskip0pt
\noindent
\begin{framed}
\footnotesize
\noindent
\vskip0pt\noindent\hskip0pt
\texttt{Lemma} \textit{poly\_even\_odd} $p$ \texttt{:}
(\textit{even\_poly} $p$ $\backslash$\texttt{Po} \textquotesingle\texttt{X\^}2) 
\texttt{+} (\textit{odd\_poly} $p$ $\backslash$\texttt{Po} \textquotesingle\texttt{X\^}2)) 
\texttt{*} \textquotesingle\texttt{X} $=$ $p$.
\vskip0pt
\noindent
\end{framed}
\vskip0pt
\noindent
We are now ready to present the algorithm we have proved correct:
\vskip0pt
\noindent
\begin{framed}
\footnotesize
\noindent
\vskip0pt\noindent\hskip0pt
\texttt{Fixpoint} \textit{fft} ($n$ : \texttt{nat}) ($w$ : $R$) ($p$ : \texttt{\{poly $R$\}}) \texttt{:} \texttt{\{poly $R$\}} \texttt{:=} 
\vskip0pt\noindent\hskip10pt
  \texttt{if} $n$ \texttt{is} $n_1$\texttt{.+1} \texttt{then}
\vskip0pt\noindent\hskip20pt
  \texttt{let} $ev$ \texttt{:=} \textit{fft} $n_1$ ($w$ \texttt{\^}\texttt{+} 2) (\texttt{even\_poly} $p$) \texttt{in}
\vskip0pt\noindent\hskip20pt
  \texttt{let} $ov$ \texttt{:=} \textit{fft} $n_1$ ($w$ \texttt{\^}\texttt{+} 2) (\texttt{odd\_poly}\,\,\,\,\,$p$) \texttt{in}
\vskip0pt\noindent\hskip20pt
  \texttt{{$\backslash$}poly\_}($i$ \texttt{<} 2\,\texttt{\^}\,($n_1$\texttt{.+1}))
  \texttt{let} $j$ \texttt{:=} $i$ \texttt{\%\%} (2\,\texttt{\^}\,$n_1$) \texttt{in} $ev$\texttt{\textquotesingle\_}$j$ \texttt{+} $ov$\texttt{\textquotesingle\_} $j$ \texttt{*} $w$ \texttt{\^}\texttt{+} $i$ 
\vskip0pt\noindent\hskip10pt
  \texttt{else} ($p$\texttt{\textquotesingle\_}0)\texttt{\%:P}.
\end{framed}
\vskip0pt
\noindent
It takes a natural number $n$, a root of unity $w$ and a polynomial $p$ and 
returns a polynomial whose coefficients are the values at the interpolation 
point $w ^ i$. More formally, the correctness lemma is the following :
\noindent
\vskip0pt
\noindent
\begin{framed}
\footnotesize
\noindent
\vskip0pt\noindent\hskip0pt
\texttt{Lemma} \textit{fftE} $n$ ($w$ : $R$) $p$ : 
\vskip0pt\noindent\hskip10pt
  \texttt{size} $p$ $\le$ 2 \texttt{\^} $n$ \texttt{$\rightarrow$} (2 \texttt{\^} n)\texttt{.-primitive\_root} $w$ \texttt{$\rightarrow$}
\vskip0pt\noindent\hskip10pt
  \textit{fft} $n$ $w$ $p$ = \texttt{$\backslash$poly\_}($i$ $<$ 2 \texttt{\^} $n$) $p$\texttt{.[}$w$ \texttt{\^}\texttt{+} $i$\texttt{]}.
\end{framed}
\vskip0pt
\noindent
The proof is straightforward. It is done by induction. We have to prove the 
equality of two polynomials, so we show that their $i^{\textit{th}}$ coefficients
are equal
\begin{framed}
\footnotesize
\noindent
\vskip0pt\noindent\hskip10pt
(\textit{fft} $n$ ($w$ \texttt{\^}\texttt{+} 2) (\textit{even\_poly} $p$))\textquotesingle\_($i$
\texttt{\%\%} 2 \texttt{\^} $n$) \texttt{+}
\vskip0pt\noindent\hskip10pt
(\textit{fft} $n$ ($w$ \texttt{\^}\texttt{+} 2) (\textit{odd\_poly}\,\, $p$))\textquotesingle\_($i$
\texttt{\%\%} 2 \texttt{\^} $n$) \texttt{*} $w$ \texttt{\^}\texttt{+} $i$ $=$ $p$\texttt{.[}$w$ \texttt{\^}\texttt{+} $i$\texttt{]}
\end{framed}
\vskip0pt
\noindent
with the assumption that $w$ is a primitive root of order $2 ^ {n+1}$.
Using the induction hypothesis on the left of the equality and the decomposition lemma
\textit{odd\_even\_polyE} on the right, we get
\begin{framed}
\footnotesize
\noindent
\vskip0pt\noindent\hskip10pt
(\textit{even\_poly} $p$)\texttt{.[}($w$ \texttt{\^}\texttt{+} 2) \texttt{\^}\texttt{+} 
($i$ \texttt{\%\%} 2 \texttt{\^} $n$)\texttt{]} \texttt{+}
\vskip0pt\noindent\hskip10pt
(\textit{odd\_poly}\,\, $p$)\texttt{.[}($w$ \texttt{\^}\texttt{+} 2) \texttt{\^}\texttt{+} 
($i$ \texttt{\%\%} 2 \texttt{\^} $n$)\texttt{]} \texttt{*} $w$ 
\texttt{\^}\texttt{+} $i$ $=$
\vskip0pt\noindent\hskip10pt
(\textit{even\_poly} $p$)\texttt{.[}($w$ \texttt{\^}\texttt{+} $i$) \texttt{\^}\texttt{+} 2\texttt{]} \texttt{+} 
(\textit{odd\_poly} $p$)\texttt{.[}($w$ \texttt{\^}\texttt{+} $i$)  \texttt{\^}\texttt{+} 2\texttt{]} \texttt{*} $w$ \texttt{\^}\texttt{+} $i$
\end{framed}
\vskip0pt
\noindent
This means we are left with proving the following equality 
\begin{framed}
\footnotesize
\noindent
\vskip0pt\noindent\hskip10pt
($w$ \texttt{\^}\texttt{+} 2) \texttt{\^}\texttt{+} ($i$ \texttt{\%\%} 2 \texttt{\^} $n$) $=$
($w$ \texttt{\^}\texttt{+} $i$) \texttt{\^}\texttt{+} 2
\end{framed}
\vskip0pt
\noindent
that directly follows from the fact that $w ^ {2 ^ {n + 1}} = 1$.

Finally, we also prove an alternative version of the algorithm that more 
explicitly exhibits the data path, the so-called butterfly. 
\begin{framed}
\footnotesize
\noindent
\vskip0pt\noindent\hskip0pt
\texttt{Fixpoint} \textit{fft$_1$} $n$ $w$ $p$ \texttt{:} \texttt{\{poly} $R$\texttt{\}} \texttt{:=} 
\vskip0pt\noindent\hskip10pt
  \texttt{if} $n$ \texttt{is} $n_1$\texttt{.+1} \texttt{then}
\vskip0pt\noindent\hskip10pt
  \texttt{let} $ev$ \texttt{:=} \textit{fft$_1$} $n_1$ ($w$ \texttt{\^}\texttt{+} \texttt{2}) (\textit{even\_poly} $p$) \texttt{in}
\vskip0pt\noindent\hskip10pt
  \texttt{let} $ov$ \texttt{:=} \textit{fft$_1$} $n_1$ ($w$ \texttt{\^}\texttt{+} 2) (\textit{odd\_poly} $p$)\,\,\, \texttt{in}
\vskip0pt\noindent\hskip10pt
  \texttt{$\backslash$sum\texttt{\_}}($j$ $<$ 2 \texttt{\^} $n_1$)
\vskip0pt\noindent\hskip20pt
    (($ev$\textquotesingle\texttt{\_}$j$ \texttt{+} $ov$\textquotesingle\texttt{\_}$j$ \texttt{*} $w$ \texttt{\^}\texttt{+} $j$) \texttt{*:} \texttt{{\textquotesingle}X\^}$j$ \texttt{+}
    ($ev$\textquotesingle\texttt{\_}$j$ \texttt{-} $ov$\textquotesingle\texttt{\_}$j$ \texttt{*} $w$ \texttt{\^}\texttt{+} $j$) \texttt{*:} \texttt{{\textquotesingle}X\^}($j$ \texttt{+} 2 \texttt{\^} $n_1$)) 
\vskip0pt\noindent\hskip10pt
  \texttt{else} ($p$\textquotesingle\texttt{\_}0)\texttt{\%:P}.
\end{framed}
\vskip0pt
\noindent
It is straightforward to prove that both recursive versions compute the same 
thing.
\begin{framed}
\footnotesize
\noindent
\vskip0pt\noindent\hskip0pt
\texttt{Lemma} \textit{fft$_1$E} $n$ ($w$ \texttt{:} $R$) $p$ \texttt{:} (2 \texttt{\^} $n$)\texttt{.-primitive\_root} $w$ \texttt{$\rightarrow$} \textit{fft$_1$} $n$ $w$ $p$ $=$ \textit{fft} $n$ $w$ $p$\texttt{.}
\end{framed}
\vskip0pt
\noindent

\subsection*{The iterative algorithm}

In our formalisation, we are going to derive an iterative version from
the recursive in a very straightforward way. Let us explain it on an example 
with a polynomial of degree 7 ($2^3 -1$). The depth of the recursion is 3
and the binary tree  of the recursive calls looks like:
\begin{framed}
\footnotesize
\noindent
\begin{center}
\begin{tikzpicture}[level distance=1.5cm,
  level 1/.style={sibling distance=6cm},
  level 2/.style={sibling distance=3cm},
  level 3/.style={sibling distance=1.5cm},
  ]
  \node {$a_0 + a_1 X + a_2 X^2 + a_3 X^3 + a_4 X^4 + a_5 X^5 + a_6 X^6 + a_7 X^7$}
    child {node {$a_0  + a_2 X + a_4 X^2 + a_6 X^3$}
      child {node {$a_0  + a_4 X$}
        child {node {$a_0$}}
        child {node {$a_4$}}
      }
      child {node {$a_2 + a_6 X$}
        child {node {$a_2$}}
        child {node {$a_6$}}
      }
    }
    child {node {$a_1  + a_3 X + a_5 X^2 + a_7 X^3$}
      child {node {$a_1  + a_5 X$}
        child {node {$a_1$}}
        child {node {$a_5$}}
      }
      child {node {$a_3  + a_7 X$}
        child {node {$a_3$}}
        child {node {$a_7$}}
      }
    };
\end{tikzpicture}
\end{center}
\end{framed}
\vskip0pt
\noindent
The idea is to put all the results at depth $i$ in a single polynomial.
Here, at depth 3 it is a polynomial containing 8 sub polynomials of degree 0,
at depth 2, 4 polynomials of degree 1, at depth 1, 2 polynomials of degree 3
and finally one polynomial of degree 7.
The final result is build bottom up. Initially we start with the polynomials
that contains all the leaves. Then, one step of the iteration simply take 
the results at depth $i$ and returns the results at depth $i-1$.

Let us first concentrate on the initial value, the values of all the leaves.
If we remember that we use an even/odd partition, putting the even part on 
the left and the odd part on the right, this means that if we look 
at the binary representation, the bits in reverse order from the right to the left 
gives us the sorting order. If we take our example, we have to sort 
$$[0; 1; 2 ;3 ;4; 5; 6; 7]$$
With their binary representation, it gives 
$$[0 \leadsto 000; 1  \leadsto 001; 2 \leadsto 010 \leadsto ;3 \leadsto 011 ;4 \leadsto 100; 5 \leadsto 101; 6 \leadsto 110; 7 \leadsto 111]$$
Reversing them, we get 
$$[0 \leadsto 000; 1 \leadsto 100; 2 \leadsto 010; 3 \leadsto 110;4 \leadsto 001; 5 \leadsto 101; 6 \leadsto 011; 7 \leadsto 111]$$
Translating them back to natural numbers, we have
$$[0 \leadsto 0\,\,\,\,\,\,\,\,; 1 \leadsto 4\,\,\,\,\,\,; 2 \leadsto 2\,\,\,\,\,\,;3 \leadsto 6\,\,\,\,\,\,;4 \leadsto 1\,\,\,\,\,\,; 5 \leadsto 5\,\,\,\,\,\,; 6 \leadsto 3\,\,\,\,\,\,; 7 \leadsto 7\,\,\,\,\,\,]$$

To build this initial polynomial, we first define \textit{digitn} $b$ $n$ $m$
that computes the $m^\textit{th}$ digit of $n$ in base $b$. We then use it to 
reverse a number :  \textit{rdigitn} $b$ $n$ $m$ reverses the $n$ first bits in 
base $b$ of $m$. Finally, the initial polynomial with $2^n$ terms is created 
by \textit{reverse\_poly} $n$ $p$ using an appropriate permutation of the 
coefficient of $p$. 
\begin{framed}
\footnotesize
\noindent
\texttt{Definition} \textit{digitn} $b$ $n$ $m$ \texttt{:=} ($n$ \texttt{\%/} $b$ \texttt{\^} $m$) \texttt{\%\%} $b$.
\vskip0pt\noindent\hskip0pt
\texttt{Definition} \textit{rdigitn} $b$ $n$ $m$ \texttt{:=} \texttt{$\backslash$sum\_}($i$ $<$ $n$) \textit{digitn} $b$ $m$ ($n$\texttt{.-1} $-$ $i$) \texttt{*} $b$ \texttt{\^} $i$.
\vskip0pt\noindent\hskip0pt
\texttt{Definition} \textit{reverse\_poly} $n$ ($p$ \texttt{:} \texttt{\{poly} $R$\texttt{\}}) \texttt{:=}
  \texttt{$\backslash$poly\_}($i$ $<$ 2 \texttt{\^} $n$) $p$\textquotesingle\texttt{\_}(\textit{rdigitn} 2 $n$ $i$).
\end{framed}
\vskip0pt
\noindent
On our example, \textit{reverse\_poly} $3$ $p$ returns a polynomial with $2^3$ terms.
$$p_0 + p_4 X + p_2 X^2  + p_6 X^3 + p_1 X^4  + p_5 X^5 + p_3 X^6 + p_7 X^7$$

Now, we want to express that after each step of the iteration we get all the 
results at depth $i$. In the recursion, the polynomial is split in two 
using the even and odd part, but the results are glued together using the left 
to right concatenation. So, we need some operations to get the low terms or 
the high term of a polynomial.
\begin{framed}
\footnotesize
\noindent
\vskip0pt\noindent\hskip0pt
\texttt{Definition} \textit{take\_poly} $m$ ($p$ \texttt{:} \texttt{\{poly} $R$\texttt{\}}) \texttt{:=} \texttt{$\backslash$poly\_}($i$ $<$ $m$) $p${\textquotesingle}\texttt{\_}$i$.
\vskip0pt\noindent\hskip0pt
\texttt{Definition} \textit{drop\_poly} $m$ ($p$ \texttt{:} \texttt{\{poly} $R$\texttt{\}}) \texttt{:=} \texttt{$\backslash$poly\_}($i$ $<$ \texttt{size} $p$ \texttt{-} $m$) $p$\textquotesingle\texttt{\_}($i$ \texttt{+} $m$).
\vskip0pt\noindent\hskip0pt
\texttt{Lemma} \textit{poly\_take\_drop} $m$ $p$ \texttt{:}
 \textit{take\_poly} $m$ $p$ \texttt{+} \textit{drop\_poly} $m$ $p$ \texttt{*} \texttt{{\textquotesingle}X\^}$m$ $=$ $p$.
\end{framed}
\vskip0pt
\noindent
\textit{take\_poly} $m$ $p$ returns the polynomial that has the $m$ low terms of
$p$ while \textit{drop\_poly} $m$ $p$ returns the polynomial with the high terms of $p$
skipping the $m$ low terms. 

We want to express that we have in a polynomial $q$ all the results of calling 
the recursive algorithm $p$ cutting at depth $n$, so every leaf is a 
call of \textit{fft$_1$} with the appropriate part of the polynomial $p$.
This is done using a recursive predicate. If $n$ is not zero, we split 
$p$ in two with even and odd part and $q$ with left part and right part.
If $n$ is zero, we are at a leaf so the result must be a call to \textit{fft$_1$}. 
\begin{framed}
\footnotesize
\noindent
\texttt{Fixpoint} \textit{all\_results\_fft$_1$} $n$ $m$ $w$ $p$ $q$ \texttt{:=}
\vskip0pt\noindent\hskip10pt
  \texttt{if} $n$ \texttt{is} $n_1$\texttt{.+1} \texttt{then} 
\vskip0pt\noindent\hskip10pt
  \textit{all\_results\_fft$_1$} $n_1$ $m$ $w$ (\textit{even\_poly} $p$) (\textit{take\_poly} (2 \texttt{\^} ($m$ \texttt{+} $n_1$)) $q$) $\land$
\vskip0pt\noindent\hskip10pt 
  \textit{all\_results\_fft$_1$} $n_1$ $m$ $w$ (\textit{odd\_poly} $p$) \,\,(\textit{drop\_poly} (2 \texttt{\^} ($m$ \texttt{+} $n_1$)) $q$)
\vskip0pt\noindent\hskip10pt
  \texttt{else} $q$ \texttt{=} \textit{fft$_1$} $m$ $w$ $p$.
\end{framed}
\vskip0pt
\noindent
For the initial polynomial, we prove that, given a polynomial $p$ of size less than 
$2^n$ the reverse polynomial has all the results of \textit{fft$_1$} $0$ at depth $n$
\begin{framed}
\footnotesize
\noindent
\vskip0pt\noindent\hskip0pt
\texttt{Lemma} \textit{all\_results\_fft$_1$\_reverse\_poly} $p$ $n$ $w$ \texttt{:}
\vskip0pt\noindent\hskip10pt
  \texttt{size} $p$ $\le$ 2 \texttt{\^} $n$ $\rightarrow$ \textit{all\_results\_fft$_1$} $n$ $0$ $w$ $p$ (\textit{reverse\_poly} $n$ $p$).
\end{framed}
\vskip0pt
\noindent
This lemma is proved by induction on $n$.  The key lemma for proving it is 
the fact \textit{reverse\_poly} decomposes nicely using odd and even parts.
\begin{framed}
\footnotesize
\noindent
\vskip0pt\noindent\hskip0pt
\texttt{Lemma} \textit{reverse\_polyS} $n$ $p$ \texttt{:} 
\vskip0pt\noindent\hskip10pt
  \textit{reverse\_poly} $n$\texttt{.+1} $p$ \texttt{=} 
\vskip0pt\noindent\hskip10pt
  \textit{reverse\_poly} $n$ (\textit{even\_poly} $p$) \texttt{+} \textit{reverse\_poly} $n$ (\textit{odd\_poly} $p$) \texttt{*} \textquotesingle\texttt{X}\texttt{\^}(2 \texttt{\^} $n$).
\end{framed}
\vskip0pt
\noindent
Now, we can define a step of the iteration. If we are at depth $m$, 
we have $2 ^ {m+1}$ results of size $2^n$
\begin{framed}
\footnotesize
\noindent
\vskip0pt\noindent\hskip0pt
\texttt{Definition} \textit{step} $m$ $n$ $w$ ($p$ \texttt{:} \texttt{\{poly} $R$\texttt{\}}) \texttt{:=}
\vskip0pt\noindent\hskip0pt
  $\backslash$\texttt{sum\_}($l$ $<$ 2 \texttt{\^} $m$)
\vskip0pt\noindent\hskip10pt
  \texttt{let} \textit{ev} \texttt{:=} \texttt{$\backslash$poly\_}($i$ $<$ 2 \texttt{\^} $n$) $p$\texttt{\textquotesingle{\_}}($i$ \texttt{+} $l$ \texttt{*} 2 \texttt{\^} $n$\texttt{.+1}) \texttt{in}
\vskip0pt\noindent\hskip10pt
  \texttt{let} \textit{ov} \texttt{:=} \texttt{$\backslash$poly\_}($i$ $<$ 2 \texttt{\^} $n$) $p$\texttt{\textquotesingle{\_}}($i$ \texttt{+} $l$ \texttt{*} 2 \texttt{\^} $n$\texttt{.+1} \texttt{+} 2 \texttt{\^} $n$) \texttt{in}
\vskip0pt\noindent\hskip20pt
    \texttt{$\backslash$sum\_}($j$ $<$ 2 \texttt{\^} $n$)
\vskip0pt\noindent\hskip30pt
      ((\textit{ev}\texttt{\textquotesingle{\_}} $j$ \texttt{+} \textit{ov}\texttt{\textquotesingle{\_}} $j$ \texttt{*} $w$ \texttt{\^}\texttt{+} $j$) \texttt{*:} \texttt{\textquotesingle{X}\^}($j$ \texttt{+} $l$ \texttt{*} 2 \texttt{\^} $n$\texttt{.+1}) \texttt{+}
\vskip0pt\noindent\hskip30pt
      \,\,(\textit{ev}\texttt{\textquotesingle{\_}} $j$ \texttt{-} \textit{ov}\texttt{\textquotesingle{\_}} $j$ \texttt{*} $w$ \texttt{\^}\texttt{+} $j$) \texttt{*:} \texttt{\textquotesingle{X}\^}($j$ \texttt{+} $l$ \texttt{*} 2 \texttt{\^} $n$\texttt{.+1} \texttt{+} 2 \texttt{\^} $n$)).
\end{framed}
\vskip0pt
\noindent
and the correctness is that applying a step decreases the depth $m$ while increasing
the size $n$.
\begin{framed}
\footnotesize
\noindent
\vskip0pt\noindent\hskip0pt
\texttt{Lemma} \textit{all\_results\_fft$_1$\_step} $m$ $n$ $w$ ($p$ $q$ \texttt{:} \texttt{\{poly} $R$\texttt{\}}) \texttt{:}
\vskip0pt\noindent\hskip10pt
  \texttt{size} $p$ $\le$ 2 \texttt{\^} ($m$ \texttt{+} $n$)\texttt{.+1} $\rightarrow$
\vskip0pt\noindent\hskip10pt
  \texttt{size} $q$ $\le$ 2 \texttt{\^} ($m$ \texttt{+} $n$)\texttt{.+1} $\rightarrow$
\vskip0pt\noindent\hskip10pt
  \textit{all\_results\_fft$_1$} $m$\texttt{.+1} $n$ ($w$ \texttt{\^}\texttt{+} 2) $p$ $q$ $\rightarrow$
\vskip0pt\noindent\hskip10pt
  \textit{all\_results\_fft$_1$} $m$ $n$\texttt{.+1} $w$ $p$ (\textit{step} $m$ $n$ $w$ $q$).
\end{framed}
\vskip0pt
\noindent
Again, the key lemmas are the ones that show that \textit{step} behaves well 
with \textit{take\_poly} and \textit{drop\_poly}.
\begin{framed}
\footnotesize
\noindent
\vskip0pt\noindent\hskip0pt
\texttt{Lemma} \textit{take\_step} $m$ $n$ $w$ ($p$ \texttt{:} \texttt{\{poly} $R$\texttt{\}}) \texttt{:}
\vskip0pt\noindent\hskip10pt
  \texttt{size} $p$ $\le$ 2 \texttt{\^} ($m$ \texttt{+} $n$)\texttt{.+2} $\rightarrow$
\vskip0pt\noindent\hskip10pt
  \textit{take\_poly} (2 \texttt{\^} ($m$ \texttt{+} $n$)\texttt{.+1}) (\textit{step} $m$\texttt{.+1} $n$ $w$ $p$) \texttt{=}
\vskip0pt\noindent\hskip10pt
  \textit{step} $m$ $n$ $w$ (\textit{take\_poly} (2 \texttt{\^} ($m$ \texttt{+} $n$)\texttt{.+1}) $p$).
\vskip0pt\noindent\hskip0pt
\texttt{Lemma} \textit{drop\_step} $m$ $n$ $w$ ($p$ \texttt{:} \texttt{\{poly} $R$\texttt{\}}) \texttt{:}
\vskip0pt\noindent\hskip10pt
  \texttt{size} $p$ $\le$ 2 \texttt{\^} ($m$ \texttt{+} $n$)\texttt{.+2} $\rightarrow$
\vskip0pt\noindent\hskip10pt
  \textit{drop\_poly} (2 \texttt{\^} ($m$ \texttt{+} $n$)\texttt{.+1}) (\textit{step} $m$\texttt{.+1} $n$ $w$ $p$) \texttt{=}
\vskip0pt\noindent\hskip10pt
  \textit{step} $m$ $n$ $w$ (\textit{drop\_poly} (2 \texttt{\^} ($m$ \texttt{+} $n$)\texttt{.+1}) $p$).
\end{framed}
\vskip0pt
\noindent
Now, we code the iteration of \textit{step}, it is straightforward to prove that 
we get the same result as the recursive algorithm.
\begin{framed}
\footnotesize
\noindent
\vskip0pt\noindent\hskip0pt
\texttt{Fixpoint} \textit{istep\_aux} $m$ $n$ $w$ $p$ \texttt{:=}
\vskip0pt\noindent\hskip10pt
  \texttt{if} $m$ \texttt{is} $m_1$\texttt{.+1} \texttt{then} \textit{istep\_aux} $m_1$ $n$\texttt{.+1} $w$ (\textit{step} $m_1$ $n$ ($w$ \texttt{\^}\texttt{+} (2 \texttt{\^} $m_1$)) $p$) \texttt{else} $p$.
\vskip0pt\noindent\hskip0pt
\texttt{Definition} \textit{istep} $n$ $w$ $p$ \texttt{:=} \textit{istep\_aux} $n$ $0$ $w$ (\textit{reverse\_poly} $n$ $p$).
\vskip0pt\noindent\hskip0pt
\texttt{Lemma} \textit{istep\_fft$_1$} $n$ $p$ $w$ \texttt{:} \texttt{size} $p$ $\le$ 2 \texttt{\^} $n$ $\rightarrow$ \textit{istep} $n$ $w$ $p$ \texttt{=} \textit{fft$_1$} $n$ $w$ $p$.
\end{framed}
\vskip0pt
\noindent
Similarly we prove the correctness of an alternative and more direct iterative 
version.
\begin{framed}
\footnotesize
\noindent
\vskip0pt\noindent\hskip0pt
\texttt{Definition} \textit{step$_1$} $m$ $n$ $w$ ($p$ \texttt{:} \texttt{\{poly} $R$\texttt{\}}) \texttt{:=}
\vskip0pt\noindent\hskip10pt
  \texttt{$\backslash$poly\_}($i$ $<$ 2 \texttt{\^} ($m$ \texttt{+} $n$)\texttt{.+1})
\vskip0pt\noindent\hskip10pt
    \texttt{let} $j$ \texttt{:=} $i$ \texttt{\%\%} 2 \texttt{\^} $n$\texttt{.+1} \texttt{in}
\vskip0pt\noindent\hskip10pt
    \texttt{if} $j$ $<$ 2 \texttt{\^} $n$ \texttt{then} 
\vskip0pt\noindent\hskip20pt
      $p$\texttt{\textquotesingle{\_}}$i$ \texttt{+} $p$\texttt{\textquotesingle{\_}}($i$ \texttt{+} 2 \texttt{\^} $n$) \texttt{*} $w$ \texttt{\^}\texttt{+} $j$
\vskip0pt\noindent\hskip10pt
    \texttt{else} 
\vskip0pt\noindent\hskip20pt
      $p$\texttt{\textquotesingle{\_}}($i$ \texttt{-} 2 \texttt{\^} $n$) \texttt{-} $p$\texttt{\textquotesingle{\_}}$i$ \texttt{*} $w$ \texttt{\^}\texttt{+} ($j$ \texttt{-} 2 \texttt{\^} $n$).
\vskip0pt\noindent\hskip0pt
\texttt{Lemma} \textit{step$_1$E} $m$ $n$ $w$ $p$ \texttt{:} \textit{step$_1$} $m$ $n$ $w$ $p$ \texttt{=} \textit{step} $m$ $n$ $w$ $p$.
\end{framed}
\vskip0pt
\noindent

\subsection*{The inverse algorithm}

We have seen how to go from polynomial to interpolation. What about the other 
way around. In fact, we can use the same algorithm. From the direct direction 
$R$ was an integral ring. Here, we need a field $F$. The notation for inverse 
is $x$\texttt{\^}\texttt{-1}. The idea of the inverse algorithm is to use $1/w$ 
instead of $w$.
\begin{framed}
\footnotesize
\noindent
\vskip0pt\noindent\hskip0pt
\texttt{Definition} \textit{ifft} $n$ $w$ $p$ : \texttt{\{poly} $F$\texttt{\}} \texttt{:=} (2 \texttt{\^} $n$)\texttt{\%:R} \texttt{\^}\texttt{-1}\texttt{\%}\texttt{:P} \texttt{*} (\textit{fft} $n$ $w$\texttt{\^}\texttt{-1} $p$).
\end{framed}
\vskip0pt
\noindent
where $n$\texttt{\%}\texttt{:R} is the coercion for natural number $n$ into $F$.
Its correctness follows if the characteristic of $F$ is not 2-.
\begin{framed}
\footnotesize
\noindent
\vskip0pt\noindent\hskip0pt
\texttt{Lemma} \textit{fftK} $n$ ($w$ : $F$) $p$ \texttt{:} 
\vskip0pt\noindent\hskip10pt
  2\texttt{\%:R} \texttt{!=} 0 $\rightarrow$ \texttt{size} $p$ $\le$ 2 \texttt{\^} $n$ $\rightarrow$ (2 \texttt{\^} $n$)\texttt{.-primitive\_root} $w$ $\rightarrow$
\vskip0pt\noindent\hskip10pt
  \textit{ifft} $n$ $w$ (\textit{fft} $n$ $w$ $p$) = $p$.
\end{framed}
\vskip0pt
\noindent

\section{Deriving a tactic}

One application of Fourier algorithm is to implement polynomial multiplication.
If we take two polynomials $p$ $q$ of degree $n$,
a naive way of multiplying them requires a quadratic number $n^2$ of 
multiplications, i.e. we 
perform the multiplication of each term of $p$ (there are at most $n$ of them)
by a term of $q$ (again $n$ of those). It is possible to get a faster algorithm
using a fast Fourier transform. The idea is to take an alternative representation 
for polynomials : a polynomial $p$ of degree $n$ is uniquely defined
by its value $p(x_1),..., p(x_n)$ on $n$ distinct points $x_1,\dots x_n$.
This representation is what is called polynomial interpolation and the 
$x_i$s are called the interpolation points.
A direct way to effectively construct a polynomial $p$ from its evaluation
 $p(x_1),..., p(x_n)$ is by using Lagrange polynomial $L_{n,k} = \Pi_{i \neq k}
 (x - x_i) / (x_k - x_i)$. It is easy to check that $L_{n,k}(x_j) = 1$ if 
 $j = k$ or 0 otherwise.
Then, we have 
$p = \sum_{1 \le i \le k} p(x_i) L_{n,i}$.

Computing the multiplication 
in the interpolation world is easy. The polynomial $pq$ is of degree at most $2n$
so if we have the values of $p$ and $q$ at $2n$ points $x_1,\dots, x_{2n}$, we simply need 
to perform $2n$ multiplication to get $p(x_1)q(x_1),..., p(x_{2n})q(x_{2n})$.
Of course, we need algorithms to move from the usual polynomial representation
to the interpolation one and back. This is what the fast Fourier transform gives us.
The trick is that we are going to evaluate the polynomials at very specific 
points of interpolation (the $x_i$s are roots of unit). What makes 
it work is that, for those points, the potential $n^2$ values of the 
different $(x_i) ^ j$ ($1 \le i, j \le n$) consist of only $n$ distinct values, 
so computing the $p(x_1),..., p(x_n)$ can be done very efficiently in 
$n \textit{log}(n)$.

Formally this gives
\begin{framed}
\footnotesize
\noindent
\vskip0pt\noindent\hskip0pt
\texttt{Definition} \textit{poly\_mul} ($p_1$ $p_2$ \texttt{:} \texttt{\{poly} $R$\textit{\}}) \texttt{:=} 
\vskip0pt\noindent\hskip10pt
  (\texttt{$\backslash$poly\_}($i$ $<$ \textit{minn} (\texttt{size} $p_1$) (\texttt{size} $p_2$)) ($p_1$\texttt{\textquotesingle\_}$i$ $*$ $p_2$\texttt{\textquotesingle\_}$i$)).
\vskip0pt\noindent\hskip0pt
\texttt{Definition} \textit{fft\_mul} $n$ $w$ ($p_1$ $p_2$ \texttt{:} \texttt{\{poly} $F$\texttt{\}}) \texttt{:=} 
\vskip0pt\noindent\hskip10pt
  \texttt{let} $p_1'$ \texttt{:=}  \textit{fft} $n$ $w$ $p_1$ \texttt{in}
\vskip0pt\noindent\hskip10pt
  \texttt{let} $p_2'$ \texttt{:=}  \textit{fft} $n$ $w$ $p_2$ \texttt{in}
\vskip0pt\noindent\hskip10pt
  \textit{ifft} $n$ $w$ (\textit{poly\_mul} $p_1'$ $p_2'$).
\vskip0pt\noindent\hskip0pt
\end{framed}
\vskip0pt
\noindent
with its associated correctness theorem.
\begin{framed}
\footnotesize
\noindent
\vskip0pt\noindent\hskip0pt
\texttt{Lemma} \textit{fft\_mul\_correct} $n$ ($p_1$ $p_2$ \texttt{:} \texttt{\{poly} $F$\texttt{\}}) $w$ \texttt{:}
\vskip0pt\noindent\hskip10pt
  $2$\texttt{\%:R} \texttt{!=} $0$ $\limp$ (\texttt{size} $p_1$ $+$ \texttt{size} $p_2$)\texttt{.-1}  $\le$ $2$ \texttt{\^} $n$ $\limp$ 
\vskip0pt\noindent\hskip10pt
  ($2$ \texttt{\^} $n$)\texttt{.-primitive\_root} $w$ $\limp$
\vskip0pt\noindent\hskip10pt
  \textit{fft\_mul} $n$ $w$ $p_1$ $p_2$ $=$ $p_1$ $*$ $p_2$.
\vskip0pt\noindent\hskip0pt
\end{framed}
\vskip0pt
\noindent

In {\Rocq} there is a reflexive tactic \textit{ring} tactic that solves 
equations in a commutative ring structure~\cite{ring}. For example, let us 
consider the equation $(a - b) * (a + b) = a^2 - b^2$. The tactic first subtracts the 
two components of the equation $(a - b) * (a + b) - (a^2 - b^2)$ 
and computes a normal form by developing products and simplifying terms.
If the normal form is $0$, the initial equation holds. We have developed 
an alternative tactic that works only for univariate polynomials over an arbitrary ring.
It can then be used to prove equations over univariate polynomials.  
It works similarly by computing a norm form but using the Fourier transform when developing
products. More precisely, we use the following datatype to represent object on which
to compute the normal form.

\begin{framed}
\footnotesize
\noindent
\vskip0pt\noindent\hskip0pt
\texttt{Inductive} \textit{pexpr} \texttt{:=} 
\vskip0pt\noindent\hskip0pt
\texttt{|} \textit{Pcons} ($a$ \texttt{:} \texttt{nat}) \texttt{|} \textit{Pxn} ($n$ \texttt{:} \texttt{nat}) \texttt{|} \textit{Px} 
\vskip0pt\noindent\hskip0pt
\texttt{|} \textit{Popp} ($p$ \texttt{:} \textit{pexpr}) \texttt{|} \textit{Padd} ($p_1$ $p_2$ \texttt{:} \textit{pexpr}) \texttt{|} \textit{Pmult} ($p_1$ $p_2$ \texttt{:} \textit{pexpr}).
\vskip0pt\noindent\hskip0pt
\end{framed}
\vskip0pt
\noindent
with the interpretation defined as follows.
\begin{framed}
\footnotesize
\noindent
\vskip0pt\noindent\hskip0pt
\texttt{Fixpoint} \textit{pexpr2poly} ($e$ : \textit{pexpr}) : \texttt{\{poly} $R$\texttt{\}} \texttt{:=}
\vskip0pt\noindent\hskip0pt
\texttt{match} $e$ \texttt{with}
\vskip0pt\noindent\hskip0pt
\texttt{|} \textit{Pcons} $a$ $\limp$ a\texttt{\%:R}
\vskip0pt\noindent\hskip0pt
\texttt{|} \textit{Pxn} $n$ $\limp$ \texttt{$\textquotesingle$X $\hat{}$ $n$} 
\vskip0pt\noindent\hskip0pt
\texttt{|} \textit{Px} $\limp$ \texttt{$\textquotesingle$X}
\vskip0pt\noindent\hskip0pt
\texttt{|} \textit{Popp} $p$ $\limp$ $-$ (\textit{pexpr2poly} $p$)
\vskip0pt\noindent\hskip0pt
\texttt{|} \textit{Padd} $p_1$ $p_2$ $\limp$ \textit{pexpr2poly} $p_1$ $+$ \textit{pexpr2poly} $p_2$
\vskip0pt\noindent\hskip0pt
\texttt{|} \textit{Pmult} $p_1$ $p_2$ $\limp$ \textit{pexpr2poly} $p_1$ $*$ $pexpr2poly$ $p_2$
\vskip0pt\noindent\hskip0pt
\texttt{end.}
\vskip0pt\noindent\hskip0pt
\end{framed}
\vskip0pt
\noindent
Note that we limit coefficients to be natural numbers. This means that we can
inject an element of \textit{pexpr} over any ring $R$ into a polynomial 
with integer coefficients.
In order to use Fourier multiplication, we need a field with primitive 
roots. So, we are actually going to perform the computation on $\mathbb{C}$.
More precisely, we use the capability of the library {\sc CoqInterval}~\cite{coqinterval}
to compute with real numbers. 
Real numbers are axiomatised in {\Rocq}. So we cannot directly compute with them as we do for example with integer numbers. 
For example, \texttt{PI} is defined as $2 p$ where
p is the value between 7/8 and 7/4 such that $\cos p = 0$. The {\sc CoqInterval} library 
provides for any real expression $e$ an interval of rational numbers $(r_1,\, r_2)$ such 
that $r_1 \le e \le r_2$. If we apply it with the constant \texttt{PI}
and the default precision, it gives the following interval
${(7074237752028440 * 2 ^{-51},\, 7074237752028441 * 2 ^{-51})}.$
For example, it can then be used to prove automatically standard 
approximations of \texttt{PI} like the following.
\begin{framed}
\footnotesize
\noindent
\vskip0pt\noindent\hskip0pt
\texttt{Lemma} \textit{pi\_naive\_approx} \texttt{:} $355 /113 - 1 / 10\, \hat{}\, 6 < \texttt{PI} < 355 /113$.
\vskip0pt\noindent\hskip0pt
\texttt{Proof}. \texttt{by} \textit{interval}. \texttt{Qed}.
\vskip0pt\noindent\hskip0pt
\end{framed}
\vskip0pt
\noindent
The idea now is to build on top of {\sc CoqInterval} the capability of
computing with $\mathbb{C}$. We use rectangular box, i.e 
a pair of intervals, one for the real part and one for the imaginary part.
For example, the pair of intervals associated to $i$ is 
$((0,\, 0),\, (1,\, 1))$. We then equip $\mathbb{C}$ with an interval arithmetic.
We can then derive a certified implementation of polynomial multiplication with coefficients in 
interval arithmetic over $\mathbb{C}$ that uses the Fourier transform.  

Now, the last step is to be able to return  
to  polynomial over $\mathbb{Z}$. Remember we want to compute the product of two polynomials 
with integer coefficients. We do the computation in an interval arithmetic over $\mathbb{C}$.
Now, each coefficient  of the resulting polynomial is in a box. We know that the result is a polynomial with integer 
coefficients. The multiplication has been proved correct, so we know that each 
box must contain the integer corresponding to the result. But if the computation is precise enough, 
each box will be such that it contains only one integer, so we can directly extract the resulting integer coefficients out of the boxes. 
Let us take an example with the product
$(\texttt{{\textquotesingle}X} - 1) * (\texttt{{\textquotesingle}X} + 1)$.
The first polynomial is represented by a list of two boxes 
$[((-1,-1), (0,0)); ((1,1), (0,0))]$. 
The second polynomial is also represented by a list of two boxes
$[((1,1), (0,0)); ((1,1), (0,0))]$. 
Running the multiplication returns a list of four boxes: \\
$[((-2050 * 2^{-11}, -4093 * 2^{-12}), (-2048 * 2^{-22}, 2048 * 2^{-21}));$\\
$\hphantom{[}((-3077 * 2^{-22},  \hphantom{-}3077 * 2^{-22}), (-3590 * 2^{-22}, 3590 * 2^{-22}));$\\ 
$\hphantom{[}(( \hphantom{-}4093 * 2^{-12},  \hphantom{-}2050 * 2^{-11}), (-2048 * 2^{-21}, 2048 * 2^{-22}));$\\
$\hphantom{[}((-3077 * 2^{-22};  \hphantom{-}3077 * 2^{-22}), (-3590 * 2^{-22}; 3590 * 2^{-22}))]$\\
It is then easy to check that
$-1$ is the only integer inside the first box,
$0$ is the only integer inside the second and forth boxes, and
$1$ is the only integer inside the third box. 
This gives us the expected result 
$\texttt{{\textquotesingle}X}\,\,\texttt{$\hat{}$}\,\,2 - 1$.

We pack the different steps (reification, injection into interval polynomial, normalisation,
extraction) in a tactic \textit{poly\_ring} that automatically proves 
equations between univariable polynomials with integer coefficients over an arbitrary ring.
\begin{framed}
\footnotesize
\noindent
\vskip0pt\noindent\hskip0pt
\texttt{Goal} \texttt{forall} $R$, $(\texttt{{\textquotesingle}X} - 1) * (\texttt{{\textquotesingle}X} + 1) = \texttt{{\textquotesingle}X}\,\,\texttt{$\hat{}$}\,\,2 - 1$ \texttt{:>} \texttt{\{poly $R$\}}
\vskip0pt\noindent\hskip0pt
\texttt{Proof}. \texttt{by} \textit{poly\_ring}. \texttt{Qed}.
\vskip0pt\noindent\hskip0pt
\end{framed}
\vskip0pt
\noindent

\section{The algorithm and floating-point numbers}

The previous section was a toy application of the Fourier transform.
Here, we are interested in more realistic applications of the Fourier transform.
A typical way of computing Fourier transform is by using floating-point 
numbers. As the computation is not exact, it is important
to quantify the error. In this section, we explain how this has been
done formally. For this, we follow the pen-and-paper proof given in~\cite{floatFourier}.
Our formalisation is based on the library {\sc Flocq}\cite{Flocq}.
In this library, the floating-point numbers are a subset of the real numbers.
There is a rounding function \texttt{RN} that is parametrized by a rounding 
mode $r$, a basis $\beta$, and a format $f$. In the following, when not stating, 
the rounding  mode is ``round to the nearest'', the basis 2, and a format 
corresponding to the IEEE 754 standard but with no upper bound on the exponent
(no overflow). This corresponds to the setting of our reference paper 
\cite{floatFourier}.      

Prior to tackling the algorithm, we first need to quantify the error made by complex multiplications.
Two types of multiplication are used one without fused multiply add ({\sc FMA}) 
and one without.
Here is the code of the basic one without {\sc FMA}.
\begin{framed}
\footnotesize
\noindent
\texttt{Definition} \textit{brent\_imult} ($z_1$ $z_2$ \texttt{:} $C$) \texttt{:} $C$ \texttt{:=} 
\vskip0pt\noindent\hskip10pt
  \texttt{let} ($x_1$, $y_1$) \texttt{:=} $z_1$ \texttt{in} 
\vskip0pt\noindent\hskip10pt
  \texttt{let} ($x_2$, $y_2$) \texttt{:=} $z_2$ \texttt{in}
\vskip0pt\noindent\hskip10pt
  (\texttt{RN} (\texttt{RN}($x_1$ $*$ $x_2$) $-$ \texttt{RN}($y_1$ $*$ $y_2$)), \texttt{RN} (\texttt{RN} ($x_1$ $*$ $y_2$) $+$ \texttt{RN} ($y_1$ $*$ $x_2$))).
\vskip0pt\noindent\hskip0pt
\end{framed}
\vskip0pt
\noindent
In {\Rocq}, complex numbers are implemented as a pair of real numbers. 
In order to multiply to complex numbers $z_1$ and $z_2$, we first extract 
their real parts and imaginary parts and then compose the result adding 
a rounding operation after each arithmetic operation. To establish a bound on
the error we follow the pen-and-paper proof given in~\cite{complexM}.
\begin{framed}
\footnotesize
\noindent
\texttt{Lemma} \textit{error\_brent\_imult} $z_1$ $z_2$ :
\vskip0pt\noindent\hskip10pt
  $2$ \texttt{\^} $5$ $*$ $u$ $\le$ $1$ $\limp$
  \texttt{Cmod} (\textit{brent\_imult} $z_1$ $z_2$ $-$ $z_1$ $*$ $z_2$) $\le$  \texttt{sqrt} $5$ $*$ $u$ $*$ \texttt{Cmod} ($z_1$ $*$ $z_2$).
\vskip0pt\noindent\hskip0pt 
\end{framed}
\vskip0pt
\noindent
where $p$ is the precision (the number of bits of the mantissa), $u$ the unit in the last place 
($u = \beta ^{1 - p} /2$), and \texttt{Cmod} the usual modulus for complex numbers.

For the multiplication using {\sc FMA}, it is derived from an algorithm given by 
Kahan to compute the 2x2 determinant~\cite{Kahan}. Here is the algorithm
that uses 2 {\sc FMA}s :
\begin{framed}
\footnotesize
\noindent
\texttt{Definition} \textit{kahan} $a$ $b$ $c$ $d$ \texttt{:=} 
\vskip0pt\noindent\hskip10pt
  \texttt{let} $w$ \texttt{:=} \texttt{RN} ($b$ $*$ $c$) \texttt{in}
\vskip0pt\noindent\hskip10pt
  \texttt{let} $e$ \texttt{:=} \texttt{RN} ($w$ $-$ $b$ $*$ $c$) \texttt{in}
\vskip0pt\noindent\hskip10pt
  \texttt{let} $f$ \texttt{:=} \texttt{RN} ($a$ $*$ $d$ $-$ $w$) \texttt{in}
\vskip0pt\noindent\hskip10pt
  \texttt{RN} ($f$ $+$ $e$).
\vskip0pt\noindent\hskip0pt
\end{framed}
\vskip0pt
\noindent
and the corresponding error as given in~\cite{Kahan}.
\begin{framed}
\footnotesize
\noindent
\texttt{Lemma} \textit{kahan\_bound} \texttt{:} 
\vskip0pt\noindent\hskip10pt
  \texttt{even} \texttt{beta} $\limp$ \texttt{Rabs} (\textit{kahan} $a$ $b$ $c$ $d$ $-$ ($a$ $*$ $d$ $-$ $b$ $*$ $c$)) $\le$ $2$ $ * $ $u$ $*$ \texttt{Rabs} ($a$ $*$ $d$ $-$ $b$ $*$ $c$).
\vskip0pt\noindent\hskip0pt
\end{framed}
\vskip0pt
\noindent
Note that this theorem is valid only if the basis $\beta$ is even.

Once we get the elementary block proved, it is straightforward to 
follow the proof given in~\cite{floatFourier}. 
\begin{framed}
\footnotesize
\noindent
\texttt{Lemma} \textit{cnorm\_floatDexact} \textit{fma} $n$ ($w$ : $C$) ($l$ : \texttt{list} $C$) :
\vskip0pt\noindent\hskip0pt
\texttt{let} \textit{omega} $k$ \texttt{:=} $u$ $+$ \texttt{gk} $n$\texttt{.-1} $w$ \textit{fma} $k$ $*$ ($1$ $+$ $u$) \texttt{in}
\vskip0pt\noindent\hskip10pt
  \texttt{size} $l$ $=$ ($2$ \texttt{\^} $n$) $\limp$ 
\vskip0pt\noindent\hskip10pt
  (\texttt{forall} $i$ : \texttt{nat}, ($i$ $<$ $2$ \texttt{\^} $n$) $\limp$ \texttt{iformat} (\texttt{get} $l$ $i$)) $limp$
\vskip0pt\noindent\hskip10pt
  ($w$ \texttt{\^} $2$ \texttt{\^} $n$\texttt{.-1}) $=$ $1$ $\limp$
\vskip0pt\noindent\hskip10pt
   (\textit{fma} $=$ \texttt{false} $\limp$ $2$ \texttt{\^} $5$ $*$ $u$ $\le$ 1) $\limp$
\vskip0pt\noindent\hskip10pt
  \texttt{cnorm} $n$ 
      (\texttt{csub} $n$ (\textit{fft\_float} \textit{fma} $n$ $w$ $l$) (\textit{fft\_exact} $n$ $w$ $l$)) $\le$ 
\vskip0pt\noindent\hskip10pt
    \texttt{cnorm} $n$ (\textit{fft\_exact} $n$ $w$ $l$) $*$ 
    ((\texttt{$\backslash$prod\_}($i$ $<$ $n$) ($1$ $+$ \textit{omega} $i$) $-$ $1$)).
\vskip0pt\noindent\hskip0pt
\end{framed}
\vskip0pt
\noindent
Our theorem is very close to the one of the pen-and-paper proof~\cite{floatFourier}.
\begin{framed}
\includegraphics[width=13cm]{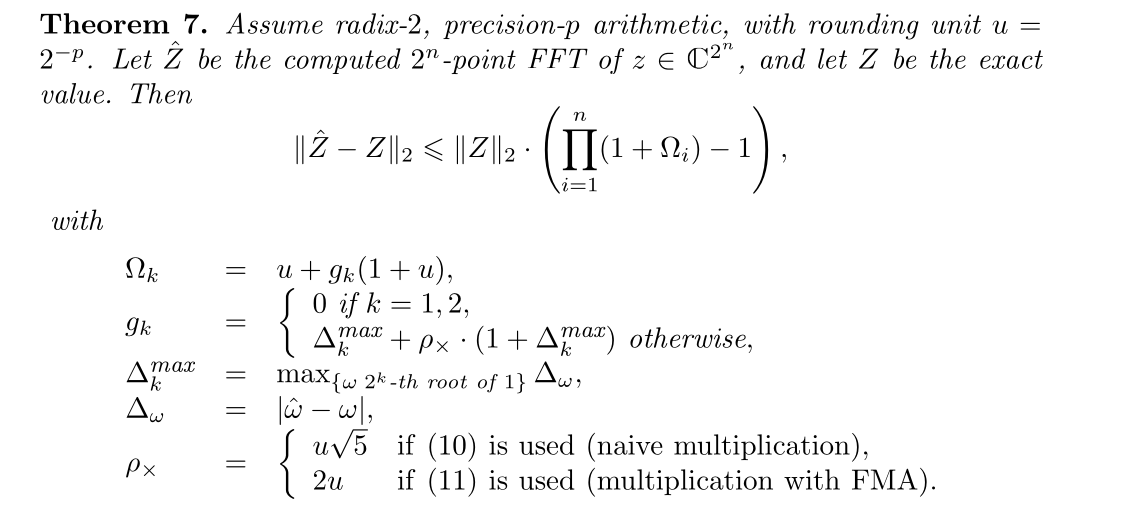}
\end{framed}
\vskip0pt
\noindent
Only the condition $2 ^ 5 * u \le 1$
is mentioned explicitly since it comes from~\cite{complexM}.
 
\section{Conclusion}

We have presented our formalisation of the Fast Fourier transform.
The complete source code is about 11000 lines. It is available at 

\url{https://github.com/thery/mathcomp-extra/blob/master/fft.v}

\noindent
We hope that what we have presented shows the versatility of what can
be done in a proof assistant. In the first part, the formalisation 
of the algorithm itself, we believe it illustrates the benefit of 
having the Mathematical component library that contains algebraic structures,
big operators and polynomials. This makes the formalisation much easier compared 
to~\cite{capretta}. The second part with the toy tactic was mostly motivated 
by our eagerness to execute our algorithm on actual values.
The tactic per se is not really of practical use. The multiplication using 
Fourier transform outperforms the naive multiplication only when the number 
of coefficients is quite high. This is very unlikely to occur while verifying equation 
over univariate polynomials. Finally, the application for floating-point numbers opens the 
opportunity to verify actual code that makes use of the FFT.

\bibliographystyle{plain}
\bibliography{Note}

\begin{thebibliography}{1}

\bibitem{Flocq}
Sylvie Boldo and Guillaume Melquiond.
\newblock {F}locq: {A} {U}nified {L}ibrary for {P}roving {F}loating-{P}oint
  {A}lgorithms in {C}oq.
\newblock In {\em 20th {IEEE} {S}ymposium on {C}omputer {A}rithmetic}, pages
  243--252. {IEEE} Computer Society, 2011.

\bibitem{complexM}
Richard~P. Brent, Colin Percival, and Paul Zimmermann.
\newblock Error bounds on complex floating-point multiplication.
\newblock {\em {M}athematics of {C}omputation}, 76(259):1469--1481, 2007.

\bibitem{floatFourier}
Nicolas Brisebarre, Mioara Joldes, Jean{-}Michel Muller, Ana{-}Maria Nanes, and
  Joris Picot.
\newblock Error analysis of some operations involved in the cooley-tukey fast
  fourier transform.
\newblock {\em {ACM} {TOMS}.}, 46(2):11:1--11:27, 2020.

\bibitem{capretta}
Venanzio Capretta.
\newblock {C}ertifying the {F}ast {F}ourier {T}ransform with {C}oq.
\newblock In {\em TPHOLs'01}, number 2152 in LNCS, pages 169--184, Edinburgh,
  Scotland, 2001.

\bibitem{coqinterval}
Paul~Geneau de~Lamarli{\`{e}}re and Guillaume Melquiond.
\newblock Coqinterval, November 2024.

\bibitem{ring}
Benjamin Gr{\'{e}}goire and Assia Mahboubi.
\newblock {P}roving {E}qualities in a {C}ommutative {R}ing {D}one {R}ight in
  {C}oq.
\newblock In {\em TPHOLs}, volume 3603 of {\em Lecture Notes in Computer
  Science}, pages 98--113. Springer, 2005.

\bibitem{Kahan}
Claude{-}Pierre Jeannerod, Nicolas Louvet, and Jean{-}Michel Muller.
\newblock Further analysis of kahan's algorithm for the accurate computation of
  2{\texttimes}2 determinants.
\newblock {\em {M}athematics of {C}omputation}, 82(284):2245--2264, 2013.

\end{thebibliography}

\end{document}